# Comparison of Access Control Approaches for Graph-Structured Data

Aya Mohamed[1,2][a], Dagmar Auer[1,2][b], Daniel Hofer[1,2][c] and Josef Küng[1,2][d]

[1]*Institute of Application-oriented Knowledge Processing, Johannes Kepler University Linz, Linz, Austria*
[2]*LIT Secure and Correct Systems Lab, Johannes Kepler University Linz, Linz, Austria*
{*aya.mohamed, dagmar.auer, daniel.hofer, josef.kueng*}@jku.at

Abstract: Access control is the enforcement of the authorization policy, which defines subjects, resources, and access rights. Graph-structured data requires advanced, flexible, and fine-grained access control due to its complex structure as sequences of alternating vertices and edges. Several research works focus on protecting property graph-structured data, enforcing fine-grained access control, and proving the feasibility and applicability of their concept. However, they differ conceptually and technically. We select works from our systematic literature review on authorization and access control for different database models in addition to recent ones. Based on defined criteria, we exclude research works with different objectives, such as no protection of graph-structured data, graph models other than the property graph, coarse-grained access control approaches, or no application in a graph datastore (i.e., no proof-of-concept implementation). The latest version of the remaining works are discussed in detail in terms of their access control approach as well as authorization policy definition and enforcement. Finally, we analyze the strengths and limitations of the selected works and provide a comparison with respect to different aspects, including the base access control model, open/closed policy, negative permission support, and datastore-independent enforcement.

## 1 INTRODUCTION

Access control ensures data security by protecting assets and private information against unauthorized access by defined subjects. It refers to the enforcement of authorization policies, which specify access rights in terms of the accessing subject, the requested resource and the performed action.

In graphs, data are structured as vertices connected by edges to represent the relationships between objects. Vertices and edges are stored as entities in graph databases, optionally including properties as key-value pairs. Fine-grained access control for graph-structured data refers to protecting vertices and edges at the attribute level. Established graph databases currently support role-based access control (RBAC), while the latest research works address relationships in graph-structured data, including attributes on vertices and edges.

We selected the following access control approaches for graph-structured data from our systematic literature review (SLR) on access control for different database models including further recent related works. Rizvi and Fong (2018) extended the relationship-based access control model (ReBAC) to support attributes using their own authorization policy specification and query evaluation algorithm. Mohamed et al. (2023a) presented a flexible, fine-grained authorization policy specification for graph-structured data and datastore-independent enforcement. Furthermore, Hofer et al. (2023a,b) proposed an approach to enforce fine-grained access control by rewriting Cypher queries using an Abstract Syntax Tree (AST). Bereksi Reguig et al. (2024) recently introduced an approach based on the Neo4j access control model to apply attribute-based access control (ABAC).

In this paper, we study these works on the conceptual and technical level through answering the following research questions:

**RQ1** What are the current access control approaches in the context of graph-structured data?

**RQ2** How are the authorization policies defined and enforced in each work?

[a]  https://orcid.org/0000-0001-8972-625
[b]  https://orcid.org/0000-0001-5094-2248
[c]  https://orcid.org/0000-0003-0310-1942
[d]  https://orcid.org/0000-0002-9858-837X

**RQ3** How are the feasibility and applicability of the proposed approaches proved?

**RQ4** What are the strengths and limitations for each approach?

The rest of the paper is structured as follows. In Section 2, we describe our method for selecting related works and the reasons for filtering out the rest. In Section 3 to 6, we discuss (1) the access control model and the conceptual approach, (2) the definition of the authorization policy including requests, (3) the policy processing and enforcement, and (4) implementation details and the evaluation for each of the selected works respectively. We compare and analyze the different approaches in Section 7. The paper concludes with a summary and an outlook on future work in Section 8.

## 2 SELECTION METHOD

In Mohamed et al. (2023b), we presented an SLR of authorization and access control for different database models, where we identified a list of works related to the graph database model. We additionally include recent related works as listed in Table 1. For each work, we indicate whether or not it is included according to our selection criteria: (1) protecting property graph-structured data, (2) fine-grained authorization policy, and (3) applied in graph datastores.

Morgado et al. (2018) present a model-based approach using metadata with authorization rules to control access in applications that use graph databases. It provides a security model for access control using a predefined schema for the graph vertices. This approach is not fine-grained, as it only considers protection on vertex level. Furthermore, it has a different objective, i.e., authorization policy representation in terms of users associated to permissions to operate on resources. The model acts as an intermediate plugin between the system and the database.

Bertolissi et al. (2019) introduce an access control framework for secure data fusion in cooperative systems. Authorization policies are represented in a provenance graph, which relates artefacts (vertices) to depict their provenance through a system. They provide the formal semantics of the proposed language and use the extensible access control markup language (XACML)[1] to demonstrate how provenance-based access constraints can be accommodated in ABAC policies. Jin and Kaja (2019) use graphs to represent the XACML authorization policy language model. An authorization policy graph is constructed by parsing the XACML policy files in addition to handling duplicates and conflicts to refine the results before generating the corresponding Cypher query statements. Accordingly, XACML requests are processed as a query to the Neo4j graph database, which stores the policy graph. However, both approaches do not consider protecting graph-structured data.

Chabin et al. (2021) propose an access control system for graph-based models with schema constraints, authorization rules to protect the data and user context rules. To enforce access control constraints, an architecture with modules for rewriting, planning and executing queries in parallel is provided. Roles can have fine-grained access rules on vertices, and clearance levels can be associated to roles and resources. However, each user has only a single role in this model. Furthermore, only type and direction can be specified for edges in the authorization rules, i.e., edge attributes are not considered. We did not consider this work in our comparison because the approach is not designed to deal with property graphs in terms of vertices and edges having attributes, but rather the resource description framework (RDF) graph model, which stores data as triples.

Valzelli. et al. (2020) introduce a property graph model, which combines discretionary access control (DAC), mandatory access control (MAC) and RBAC to protect knowledge graphs. Positive and negative authorizations can be specified using authorization edges between subjects and resources. The RDF concepts are systematically translated into a property graph query. The concept is implemented using Tinkerpop[2], as it is compatible with the most common graph datastores. We excluded RBAC approaches because this access control model is not fine-grained and is already supported in many graph databases, such as Neo4j[3], ArangoDB[4] and Azure Cosmos DB[5]. Therefore, Valzelli. et al. (2020) is not selected for the detailed study.

Clark et al. (2022) define ReLOG as a formal language and a unified framework to encode ReBAC policies, which can naturally be viewed as graph queries over graph databases. Although the work extended ReBAC to be fine-grained using graph patterns, we excluded it because only the theoretical feasibility of the approach has been investigated. Neither

---

[1] http://docs.oasis-open.org/xacml/3.0/xacml-3.0-core-spec-os-en.html

[2] https://tinkerpop.apache.org/
[3] https://neo4j.com/blog/role-based-access-control-neo4j-enterprise/
[4] https://docs.arangodb.com/stable/arangograph/security-and-access-control/
[5] https://docs.microsoft.com/en-us/azure/cosmos-db/role-based-access-control

Table 1: Related works in access control for the graph database model.

| Research works | Protect property graph | Fine-grained | Applied |
|---|---|---|---|
| Rizvi and Fong (2018) | ✓ | ✓ | ✓ |
| Morgado et al. (2018) | × | × | ✓ |
| Bertolissi et al. (2019) | × | ✓ | ✓ |
| Jin and Kaja (2019) | × | ✓ | ✓ |
| Mohamed et al. (2020) | ✓ | ✓ | ✓ |
| Valzelli. et al. (2020) | ✓ | × | ✓ |
| Chabin et al. (2021) | ×[1] | ✓[2] | ✓ |
| Clark et al. (2022) | ✓ | ✓[2] | × |
| Hofer et al. (2023a) | ✓ | ✓ | ✓ |
| Bereksi Reguig et al. (2024) | ✓ | ✓ | ✓ |

[1] Based on the RDF graph model
[2] For vertices but not edges

a concrete implementation nor demo cases are provided.

Therefore, we discuss the latest research results of the remaining works in detail, which are Rizvi and Fong (2020), Mohamed et al. (2023a), Hofer et al. (2023a) and Bereksi Reguig et al. (2024).

## 3 ACCESS CONTROL APPROACH

In general, we identified two main categories for access control approaches: permit-deny access and filtered results. The *permit-deny* is the classical access control type, where access is requested giving the required information of subjects, resources, actions, and other optional conditions. The input request is checked against the defined authorization policy and a decision is returned indicating whether the access is authorized or not. This is typically used in physical access control systems to authorize or prevent access to a building. Furthermore, it can be used to control the type of actions to be performed on a specific resource, e.g., deleting a document, viewing protected details or editing personal information.

The *filtered results* category refers to returning authorized data. In case of an unauthorized access request, the result will be empty. This kind of approach is convenient when there is no specific resource to access, but rather a list of authorized resources should be retrieved at once. It is commonly used in databases to filter the query result according to the policy constraints using views or query rewriting.

Although there are currently several works that focus on data protection in graph databases, they have different objectives and use cases. In the rest of this section, we discuss the conceptual approach for each of the selected works with respect to their access control model and approach category (i.e., permit-deny access or filtered results).

**AReBAC**. The *Attribute-supporting Relationship-Based Access Control Model* (*AReBAC*) introduced by Rizvi and Fong (2018), is an extension to *ReBAC* (Bruns et al., 2012; Cheng et al., 2012; Crampton and Sellwood, 2014). Access decisions in ReBAC are based on relationships between subjects and resources. With AReBAC also attributes on vertices and edges along these relationships are considered to provide fine-grained access control. AReBAC is considered for protecting data in property graphs, i.e., edge-labeled graphs with properties (attributes) on vertices and edges.

AReBAC has been developed in the context of Neo4j, one of the most popular graph databases and its query language Cypher. The approach relies on graph patterns, a proprietary format to specify queries and access control policies independent of a specific graph datastore and query language. Therefore, a subset of Cypher, called Nano-Cypher, is defined as an equivalent language to graph pattern.

Authorization requests are specified in a pair $(m,s)$, where $m$ is the requested method (i.e., a database query with additional information such as its category) and $s$ is the subject (i.e., the user). To appropriately apply AReBAC, database queries must be defined at the application level and invoked through API methods (Rizvi, 2020). An authorization policy is defined in a graph pattern. With a request $(m,s)$ by a user $s$, the matching policies for the user's query $m$ (i.e., the method) are identified and integrated with the query. This query rewriting-based enforcement is a filtered results access control model. It does not consider data manipulating operations such as create or delete.

**XACML4G**. The work of Mohamed et al. (2020,

2021) aims to specify and enforce fine-grained, dynamic authorizations for graph-structured data. There can be several paths from a subject to a resource, but not all of them are authorized. Accordingly, specific connections to a particular object should be denied. A subject can only access a protected resource through authorized paths considering the context of resources, as it can reveal confidential information.

Therefore, the ABAC model is extended to specify flexible graph patterns by describing a path from the subject to the resource and defining fine-grained authorization constraints for vertices and edges. Additionally, pattern-related conditions can be specified to join and compare path elements. The XACML policy language model and reference architecture are extended as a proof-of-concept for the proposed approach, which is called *XACML for graphs (XACML4G)* to consider specification and evaluation of graph patterns. XACML, and hence XACML4G, is a permit-deny access model since the result is a decision.

**GQRA**[6]. Hofer et al. (2023a,b) present a graph query rewriting-based approach (GQRA) to enforce fine-grained authorizations dynamically for Cypher queries. Insecure queries are rewritten to include authorization-specific filters before being handed over to the database. Thus, only authorized data is returned, or nothing if no policy rule is matched or an error occurs during policy evaluation.

The query rewriting algorithm uses a generated AST corresponding to the source Cypher query. Based on the defined policy, authorization-specific filters are added to the syntax tree and the query is reformulated accordingly. This approach implicitly enforces ABAC in graph databases, especially Cypher-based datastores, beyond their supported access control features.

**ABAC for Neo4j**[6]. Bereksi Reguig et al. (2024) extended the RBAC model in Neo4j to support attributes. In Neo4j, the enterprise edition supports granting and denying traverse, read and match privileges to node labels and relationship types along with their properties. However, the authorization policy rules are associated to roles and apply to all nodes or relationships with the specified label or type. Therefore, they extended this access control model to support fine-grained conditions to filter out further nodes or relationships from being traversed or read.

To answer RQ1, we not only selected works re-lated to access control for graph-structured data but also classified the different access control approaches into two main categories (i.e., permit-deny and filtered results). We additionally gave an overview of the conceptual approach in each work. Further details concerning policy definition and enforcement will be discussed in the upcoming sections.

---

[6] We defined this abbreviation to easily refer to this approach throughout our paper.

## 4 POLICY DEFINITION

In this section, we explain the authorization policy language model for the selected works along with an overview of the access request and response (if any).

**AReBAC**. Rizvi (2020) defines authorization policy in the context of AReBAC as "... a graph pattern used for specifying authorization requirements for accessing resources.". It can be considered as a graph restricted by mutual exclusion constraints and attribute requirements. AReBAC assumes an underlying open policy. An authorization policy is defined for a category of database queries, at most one policy per category. Categories are used for the request and policy matching. With graph patterns, a language for declarative authorization policy specification is provided, as an easier to use alternative to logic formulas and regular expressions (Rizvi and Fong, 2020).

The following example shows a graph pattern *GP* with its components, i.e., a set of vertices $V$, a set of edges $E$, mutual exclusion constraints on vertices $\Sigma$, a set of attributes of vertices $\Gamma_V$ and edges $\Gamma_E$, a category $c$, a mapping of selected vertices $G_{ACT}$ and the return values *Ret*. The graph pattern also contains placeholders (e.g., REQ_ID) which will be replaced during request processing and policy matching.

Listing 1: Example of a graph pattern for a policy in a hospital scenario (Rizvi, 2020).

```
V      = {requestor, patient, health_record}
E      = {patient_record(patient, health_record)}
Σ      = {}
Γ_V    = {requestor.id = REQ_ID,
          health_record.id = HR_ID}
Γ_E    = {}
c      = c_read_hr
G_ACT  = {requestor_act ↦ requestor,
          health_record_act ↦ health_record,
          patient_act ↦ patient}
Ret    = {health_record}
```

To specify an authorization policy, a Nano-Cypher query with an associated category $c$ and mapping $G_{ACT}$ can also be used instead of a graph pattern, as they are equivalent.

**XACML4G.** A preliminary policy specification is defined in a proprietary JSON format in Mohamed et al. (2020). Then, the XACML4G elements (i.e., pattern, pattern condition, policy meta, vertex, and edge) are defined using XSD and applied in XACML as a state-of-the-art authorization policy language and enforcement model for ABAC.

The meta element is an extension in the XACML policy to define vertex labels and edge types of the source graph relevant for the policy evaluation. The pattern and pattern condition elements are defined in the extended XACML rule to describe constraints and joining conditions related to path elements. The pattern is composed of a path with composite structure, consisting of a vertex, an edge, and either another vertex (i.e., the base case) or an entire path. Each vertex has an identifier, a label, its own category URI and a sequence of attributes. Edges additionally include an optional direction and a type instead of a label. For flexible patterns, minimum and maximum range for part of the path can be specified as an attribute in the edge element. The pattern condition element has the same structure like the XACML condition, but the identifier attribute corresponding to the variable for a vertex or an edge in the rule pattern has to be defined. Furthermore, the function attribute should be one of the supported functions, e.g., AND, OR, (not) equal, greater/less than (or equal), string equal case insensitive, contains and starts with.

Access request is the starting point in the XACML data flow model, which contains attributes related to the subject, resource, action and others (e.g., environmental conditions) to be checked against the policy. The request is extended to include path-related attributes and specify not only vertices as resources, but also edges. Based on the input request, the policy is matched and evaluated, then an access decision is returned: permit, deny, indeterminate (i.e., more than one policy applies or an error occurs during evaluation), or not applicable (i.e., no policy is matched).

**GQRA.** The authorization policy model defined in Hofer et al. (2023a) is influenced by XACML, such that authorization requirements are represented using a policy $P$ having a set of rules $R$. The policy describes a pattern of elements $E$ (i.e., vertices and edges), which specify the path from the subject to the resource. The matching elements can be referenced for further analysis in applicability checks (i.e. conditions $C$) and filter templates $F$, which require runtime information.

Each rule $r \in R$ references a single element $e \in E$ of the policy pattern and specifies one or more boolean combined conditions $C$. A condition $c \in C$ checks whether *filter* and *return* properties are satisfied by any element of the policy pattern. The filter status indicates whether the element is filtered according to its label/type, properties or not at all. The return refers to whether the element is included in the return clause of the query (i.e., directly or aggregated) or not. The policy has a proprietary format and is checked upon receiving a Cypher query. The rewritten query only returns authorized data.

**ABAC for Neo4j.** The policy rules in Bereksi Reguig et al. (2024) are expressed like in Neo4j with an introduced `WHERE` statement to specify one or more conditions. Each condition $C$ consists of an attribute (@$a$ or $$a$), an operator or a built-in Cypher function ($op$) and a value ($val$). The attribute belongs to an entity in the graph, user or context. Thus, the policy extension is fine-grained at the vertex and edge attributes.

Currently, conditions are only supported with the *traverse* privilege. The *read* rules can be specified without conditions to avoid circular permissions. The *match* is not directly supported, but it can be expressed by combining both privileges (i.e., traverse and read). The policy is represented using the extended grammar of the Neo4j access control model in Listing 2, but stored in an independent policy-specific graph after processing. The format of access requests is also in Cypher by calling the implemented function/procedure taking the user query as input. The response is the filtered results from the data graph.

Listing 2: Policy definition in Bereksi Reguig et al. (2024).
```
T := (Grant | Deny) Traverse on Graph G
    (Nodes | Relationships) E to R
    WHERE C
C := C AND C | C OR C | NOT C | (C) |
    @a op val | $a op val
```

We summarize the main differences concerning policy definition as follows. The approaches differ concerning their policy definitions in various aspects. The filtered results approaches, i.e., AReBAC, GQRA and ABAC for Neo4j, define their policies as a kind of query or filter template to restrict the result set. AReBAC, for example, additionally uses a category with the query for matching the query with the authorization policy and determining the nodes relevant for policy evaluation. XACML4G follows the sophisticated XACML structure extended with components to define paths in the graph. Furthermore, only XACML4G and ABAC for Neo4j support posi-tive and negative permissions. While XACML4G and ABAC for Neo4j are based on established policy formats such as XACML and the Neo4j policy definition, AReBAC and GQRA define their own proprietary formats. Furthermore, only XACML4G supports a closed policy, while all others follow an open.

## 5 POLICY ENFORCEMENT

We now present the policy enforcement approach for each of the selected works, including processing, matching and conflict resolution (if applicable).

**AReBAC**. Rizvi and Fong (2018) propose an enforcement strategy based on query rewriting which returns only authorized data. However, no Nano-Cypher queries are directly executed in Neo4j, but the evaluation relies on the internal graph pattern format and the evaluation algorithm *GP-Eval*. Whenever a user sends an authorization request (*m*,*s*), the database query is translated to a graph pattern. Then the relevant authorization policy is identified via the method's category. Categories can be organized in a category hierarchy, which defines a refinement relationship between the categories, e.g., $c_1 \geq c_2$. Accordingly, the policy for methods with category $c_1$ must be at least as restrictive as the one for methods with category $c_2$. To automatically ensure this dependency (i.e., $c_2$ is more general than $c_1$), inheritance is applied. Therefore, the superior category (e.g., $c_1$) inherits all requirements from its base category (e.g., $c_2$). Therefore, all graph patterns along the inheritance hierarchy are combined into a single graph pattern. The integrated policy and the database query are then merged resulting in a single graph pattern. Rizvi and Fong (2018) call this process *weaving* and not *query rewriting*, probably as it is performed on the graph pattern. The integrated graph pattern is then evaluated with the *GP-Eval* algorithm and only authorized data is returned. As the AReBAC approach relies on its internal graph pattern format, it is datastore-independent, at least at its core. The database is accessed during the evaluation process of the graph pattern to retrieve candidates for the result set, which is the basis for finally determining the query result.

**XACML4G**. In Mohamed et al. (2023a), the XACML conceptual components are extended to match the request path with the patterns in policy rules and evaluate the pattern conditions. To allow for a source datastore-independent policy evaluation, a subset of the overall source data is managed in an independent graph, which is called source-subset graph. This graph is optionally used to evaluate the policy without relying on a specific datastore and can be constructed from multiple data sources.

The policy administration point (PAP) is extended to process the XACML4G policy and create the source-subset graph according to the specified entities in the meta element. In the extended context handler, XACML4G requests are parsed to extract the path attributes. Then, the extended policy decision point (PDP) adds a specific condition for evaluating the pattern if it exists in the matched policy rule. Based on the pattern and its conditions, a Cypher pattern and a `WHERE` statement are dynamically generated and the query is executed to evaluate path constraints as an extension in the policy information point (PIP). Conflicts are resolved in XACML at the level of policies and rules using custom or predefined combining algorithms, such as first applicable and (ordered) deny/permit overrides.

**GQRA**. Hofer et al. (2023a,b) enforces fine-grained authorizations by rewriting Cypher queries at runtime. Firstly, the query is processed by mapping each element in the query to a corresponding one in the defined policy based on the pattern structure and the element label(s). This is done by mapping the patterns to a set of paths, each one is represented as a tuple with a start vertex, an edge and an end vertex. For isolated vertices, the end vertices along with the connecting edges are empty. This step converts the patterns of policy and query into a comparable structure for policy matching. A policy is applicable if each path tuple of the insecure query matches one of its rule patterns and satisfies the conditions of this rule.

A simplified AST is built using a subset of the Cypher grammar, excluding syntactical details as well as omitting the `CREATE` and `WITH` clauses for simplicity. It describes the semantics of the source Cypher query. Filters of the matched policy rules are added to the `WHERE` part of the tree. Nodes without labels could cause ambiguity when multiple rules are matched. Thus, an optimization approach is needed to avoid overlapping filters or an empty result, due to non-matching constraints. Since the policy model only supports positive permissions, no conflicts occur. The final tree is then translated to a Cypher statement, which is sent to Neo4j to retrieve the authorized result set.

**ABAC for Neo4j**. Bereksi Reguig et al. (2024) proposed an algorithm called *SafeCypher* to rewrite a user query according to the defined authorization policy along with the user role(s) and return a safe one

(if applicable). First, the authorization graph is constructed such that each policy component (i.e., role, privilege, permission and entity) is represented as a vertex connected by directed edges, e.g., user role to privilege (i.e., traverse or read), privilege to permission, and permission to entity (i.e., node label, relationship type and attribute).

A variable is assigned to each node and relationship in the source query and the query elements (e.g., MATCH, WHERE and RETURN statements) are extracted. In the case of a nested query, the subqueries are separately rewritten and the results are appended to the main query. Then, the corresponding access conditions are added to the WHERE statement and the attributes being returned are checked. Conflicts are resolved using the authorization policy graph using the algorithm deny-override like in Neo4j. Finally, the safe query is optimized before execution.

In summary, we answered RQ2 in Section 4 and 5 by discussing the policy definition and enforcement mechanism for the selected works on the conceptual level. In the following section, we present the technical details for each work in addition to the provided use cases (if any).

## 6 IMPLEMENTATION

In this section, we address how the proposed access control approaches are feasible and applicable (refer to RQ3). For each work, we demonstrate which frameworks and graph databases have been used in addition to where in the architecture it has been implemented, e.g., within the database or an independent layer. Furthermore, we give an overview of the evaluation methods and results.

**AReBAC**. A proof of concept implementation of AReBAC[7] including selected test cases is available on GitHub. It is published under the MIT license. The pure Java implementation uses Neo4j. However, the core concept can be considered as datastore independent since it supports the graph pattern format as the internal format for representing database queries as well as authorization policies. Nano-Cypher and graph patterns can be translated to each other, as they are semantically equivalent. However, alternative semantically equivalent Nano-Cypher queries may be generated for one graph pattern, with different evaluation performance. This would require some additional query optimization, which is not provided in this implementation. In the reverse translation direction, the graph pattern is provided with no optimization required. The implementation of AReBAC is proprietary. Graph pattern is a proprietary format for the internal representation of queries and policies. Concerning the overall architecture, access control is implemented between the application and the database.

Rizvi and Fong (2020) use the *Census-Income (KDD) Dataset* from the UCI Machine Learning Repository[8] with about 200,000 entities with 40 optional attributes to demonstrate the feasibility and performance of AReBAC. Different versions of the GP-Eval algorithm are discussed and compared with Neo4j's Cypher evaluation engine, revealing significantly better performance of the GP-Eval algorithm, especially its FC-LBJ (i.e., FC for forward checking and LBJ for the Live-End Backjumping backtracking scheme) version. In addition, the performance of semantically equivalent Nano-Cypher queries may differ, while all equivalent graph patterns perform equally.

**XACML4G**. From the technical perspective, XACML is an authorization service between the datasource and the application. Thus, XACML4G (Mohamed et al., 2023a) is considered as an independent layer since the current implementation of the approach is an extension to the language and architecture of XACML. The approach is implemented as a source-code extension to Balana[9], which is an open-source framework based on Sun's XACML implementation. The proof-of-concept prototype is implemented in Java. Furthermore, the XACML4G policies and requests are validated using the XML schema for the language extensions. The optional source-subset graph is created and stored in an embedded Neo4j database. Currently, two Java classes are implemented to import graph data for Neo4j and ArangoDB.

Upon receiving an input access request, it is parsed to extract relevant attributes for policy matching and pattern evaluation. Then, the PDP proceeds with evaluating the matched policy rules to determine the access decision, including the XACML4G pattern and its conditions. A Cypher query is generated from the pattern constraints in the policy and path attributes in the extended request. The pattern evaluation is successful if the query returns a result (i.e., true) after being executed either in the source database or the source-subset graph.

Two demo cases are presented with various access control scenarios and data models in different graph database systems: the TEAM model (Hübscher et al., 2021) in the context of a research project (15K vertices and 70K edges) and the public Airbnb dataset[10] (129K vertices and 138K edges). Paths with different lengths (1 to 5) are evaluated. In contrast to previous results, the latest XACML4G prototypical implementation showed better performance and the constant overhead due to policy processing is eliminated. The overall overhead is reduced from approximately 25 (excluding processing overhead) to 21 milliseconds.

---

[7] https://github.com/szrrizvi/arebac/

[8] https://archive.ics.uci.edu/
[9] https://github.com/wso2/balana

**GQRA**. The query rewriting approach in Hofer et al. (2023a,b) is implemented in Java as an extension to the *Neo4j Object Graph Mapper (OGM)*[11]. Hence, it runs directly on top of the database as the Neo4j OGM connects to the database via the Neo4j driver. The class *Session* in the OGM framework is extended to enforce access control constraints not only when executing Cypher queries, but also in loading, saving and deleting objects. A parser is built from the relevant subset of Cypher's grammar using ANTLR[12] and used to parse the trees of the queries.

The input is a Cypher query or a request calling the OGM methods. A function `getPaths` is implemented to match the query pattern with those of the policy. Then, authorization-related filters are added to the query before sending it to the database using the presented query rewriting algorithm and placeholders of the filter template are replaced with runtime values from the source query. The resulting secure query is executed, and the output is either the authorized resources based on the combination of filter and return status or null. Only one `MATCH` statement in the query is supported.

The prototypical implementation is applied in the context of the research project *SyMSpace*[13]. Authorization rules are defined in a configuration class and enforced in the web application to load only resources the user is authorized to view. According to the graph model's structure, permissions can be assigned directly to users or via roles. The performance of the query rewriting without database access is measured using a set of representative queries. The experiments showed that the average rewriting time is $\approx 0.2$ milliseconds, which is considered to be a negligible overhead.

**ABAC for Neo4j**. Bereksi Reguig et al. (2024) implemented their proposed concept in Java within Neo4j.

---
[10] https://insideairbnb.com/get-the-data.html
[11] https://neo4j.com/developer/neo4j-ogm/
[12] https://www.antlr.org/
[13] https://web.symspace.lcm.at

It is mentioned that their Cypher query rewriting algorithm is executed within the Neo4j browser via the `CALL` clause. The input in this case is the source query, whereas the output is the result of evaluating the rewritten query. The defined authorization policy is processed from its textual representation to an access graph and stored in a Neo4j database. User-defined procedures and functions (Neo4j, 2024) can be invoked directly from Cypher and used to extend Neo4j using customized Java code.

They evaluated their implementation with respect to efficiency and scalability using the *Stack Exchange*[14] dataset. For the efficiency evaluation, they measured the execution time of 5 queries for a fixed data graph (296K vertices and 307K edges) and varied sizes of Cypher queries and authorization policies. The same experiment is conducted for the scalability evaluation, but with varying the data size and fixing the other factors. The experiments show that introducing conditions in authorization rules has minor performance overhead. It scales well with large datasets in addition to complex Cypher queries and authorization policies.

Although all selected works provide proof-of-concept implementations, the proposed approaches are still research results. AReBAC is an independent framework with a proprietary implementation for specifying authorization policies in Nano-Cypher and enforcing them using their GP-Eval algorithm. XACML4G and GQRA are implemented as extensions to Balana and OGM, respectively, to prove the feasibility of the concepts without being limited to these frameworks. Designing the implementation based on an existing solution has the advantage that it can be directly applied when the framework is used with compatible graph datastores. Finally, ABAC for Neo4j is implemented using the extension option in Neo4j to define custom procedures, which can be invoked in Cypher and perform operations on the database directly.

## 7 DISCUSSION

After the detailed discussion of the four selected approaches, which all support fine-grained access control for graph-structured data stored in property graphs, we now compare them according to the following criteria, which we consider the most important to differentiate between these models: access control approach, base access control

---
[14] https://archive.org/details/stackexchange

Table 2: Comparison of the selected works.

| Criteria | AReBAC | XACML4G | GQRA | ABAC for Neo4j |
|---|---|---|---|---|
| Access control approach | filtered results | permit-deny access | filtered results | filtered results |
| Base access control model | ReBAC | ABAC | ABAC | RBAC |
| Open policy | ✓ | ✗ | ✓ | ✓ |
| Negative permissions | ✗ | ✓ | ✗ | ✓ |
| Runtime policy processing | ✓ | ✓ | ✓ | ✗ |
| Datastore-independent | ✓ | ✓ | ✓[1] | ✗ |

[1] Cypher-based datastores only

model, open/close policy, positive and/or negative permissions, policy processing, and datastore independence. The results are summarized in Table 2.

**Access control approach**. Most of the selected approaches rely on *filtered results* access control, which allows to perform access control close to the database layer, by reducing the result set to only permitted data for a specific user. Therefore, the same access request by different users may return different result sets, depending on their authorizations. XACML4G belongs to the classic *permit-deny* category, where the access request is checked and a decision is returned, but without data.

**Base access control model**. Although fine-grained, attribute-based access control in the context of graphs is considered in all approaches, different access control models were chosen as the basis for their extensions. AReBAC was developed in the tradition of ReBAC, as Rizvi and Fong already did research in this area before, but in the context of relational databases, without considering attributes Rizvi et al. (2015). Relationships, i.e., the edges, are an essential part of the graph model, thus, the ReBAC concepts can be easily mapped to the graph model. But, it needs extensions to support attribute-based access control as well. With XACML4G and GQRA the base model is ABAC, which allows to define flexible, fine-grained authorization policies, but does not consider the specification of restrictions on paths. Therefore, these approaches extend ABAC accordingly. The XACML4G policy structure, which relies on XACML, has additional components to define path patterns and restrictions on them and meta data needed for policy evaluation. GQRA on the other hand supports the definition of path-specific requirements in their filter templates. ABAC for Neo4j in contrast is based on RBAC like access control in the Neo4j Enterprise edition and extended to support fine-grained conditions on graph entities, the user or context.

**Open/closed policy**. For the open policy, subjects are authorized by default (Samarati and de Vimercati, 2001). All filtered results approaches, i.e., AReBAC, GQRA, and ABAC for Neo4j rely on an open policy. Hence, authorizations limit the result. XACML4G assumes a closed policy, which only allows access in case of positive permissions.

**Positive and/or negative permission support**. AReBAC and GQRA policies define filters, which specify the permitted result set. However, XACML4G and ABAC for Neo4j policies must explicitly define, if matching the policy conditions results is granting or denying access.

**Runtime policy processing**. This aspect considers if policies need some pre-processing or if they can be evaluated immediately at runtime. With ABAC for Neo4j, policies are defined in a textual form and then transformed into an access graph, which is needed for query rewriting. This access graph is built in a pre-processing step. With all other approaches, the policies can be directly evaluated with each authorization request. However, pre-processing may enhance performance also with other approaches, e.g., translating AReBAC policies defined in Nano-Cypher to graph patterns in advance.

**Datastore-independent enforcement**. The AReBAC concept by Rizvi and Fong (2018) can be considered as datastore independent, as the internal representation of queries and policies is graph patterns. Thus, any graph pattern-equivalent language could be used. However, the proof-of-concept implementation relies on Neo4j and the Cypher subset Nano-Cypher. To flexibly exchange the underlying database, the implementation would need an additional abstraction layer for the database access. The query rewriting approach in Hofer et al. (2023a) can be applied to any Cypher-based datastore, while the extended authorization policy language and enforcement model of XACML4G is dynamic, applies external authorization, and deals with property graph-compatible datastores, and is thus, even more generally applicable. In contrast, the work of Bereksi Reguig et al. (2024) is datastore specific, i.e., restricted to Neo4j. Although they overcome the limitations of the Neo4j access control model which is not fine-grained and the rules have to be statically defined in the system database, the authorizations are still linked to roles and are not flexibly assigned to users according to attribute-based criteria.

Each of the approaches has its strengths and weaknesses, and their relevance heavily depends on the specific requirements. Datastore-independent enforcement, for example, may be highly appreciated from a research perspective but rather irrelevant in a practical setting. We summarize the main results to answer RQ4 in the following. The three filtered results approaches (AReBAC, GQRA, and ABAC for Neo4j) share many of the conceptual details such as an open access control approach, but differ in the chosen basic access control model. ABAC for Neo4j supports negative permissions like the authorization model of Neo4j. Furthermore, it follows the Neo4j policy definition format unlike AReBAC and GQRA, which have proprietary formats. GQRA is restricted to Cypher-based datastores as it enforces the policies by rewriting Cypher queries, whereas AReBAC "weaves" the database query and policies specified in the proprietary graph pattern format, and is thus conceptually independent of the Cypher language. XACML4G in contrast differs in fundamental criteria, as it is a *permit-deny* access control approach with a closed policy. However, like GQRA, it is based on ABAC. Furthermore, XACML4G is datastore-independent, as it can be enforced on any property graph-compatible datastore.

## 8 CONCLUSIONS

In this paper, we aim to compare and analyze access control approaches in the context of graph-structured data. Accordingly, we selected Rizvi and Fong (2018), Mohamed et al. (2020), Hofer et al. (2023a), and Bereksi Reguig et al. (2024) from a list of related works in access control for the graph database model. The proposed access control approaches belong to either the *permit-deny access* or *filtered results* category (RQ1). The first one refers to checking whether the subject is authorized to access the requested resource or not, whereas the latter returns authorized resources (if any).

We answered RQ2 and RQ3 by discussing each work in terms of access control model, authorization policy definition, policy enforcement approach and implementation details. In AReBAC, the policy is either specified using their graph pattern representation or a Nano-Cypher statement, which is then translated to a graph pattern. An enforcement algorithm called *GP-Eval* is introduced to evaluate graph patterns with a minimized number of database accesses as well as the amount of retrieved data. The concept is proprietarily implemented as an independent intermediate layer on top of the source Neo4j database. XACML4G defines graph-specific authorization constraints on vertices and edges as a recursively structured path pattern. Conditions for comparing and joining pattern elements can be specified and edges are also considered as resources. Regarding policy enforcement, dynamic queries are generated for the matched policies, taking additional attribute values from the request, and then executed in the source database or the one storing the source-subset graph. The approach is implemented as an extension to the XACML policy language and architecture. The authorization policy in GQRA is a set of rules with paths having placeholders to be replaced with runtime values. To enforce the policy, an AST equivalent to the source query is generated for adding authorization-specific conditions, and then translated back to a Cypher query. OGM is extended to apply the proposed query rewriting approach. In ABAC for Neo4j, the policy has a textual representation based on the Neo4j access control model, but stored in a graph after processing. For the policy enforcement, conditions for each clause are appended to the query, which is optimized after resolving conflicts. A user-defined procedure is implemented in Neo4j taking the source query as input to be rewritten and executed.

For the last research question (RQ4), we defined additional criteria to distinguish between the selected works, which already satisfy our initial selection criteria (i.e., protect property graph-structured data, fine-grained, and applied in graph datastores). Only XACML4G is designed differently with respect to the *permit-deny* access control approach and closed policy, while the rest belongs to the *filtered results* category with open policy. Concerning the base access control model, XACML4G and GQRA rely on ABAC, while AReBAC and ABAC for Neo4j rely on ReBAC and RBAC respectively. XACML4G and ABAC for Neo4j support negative permissions, unlike the other works. ABAC for Neo4j is the only work that supports neither datastore-independent policy enforcement nor runtime policy processing, as the approach is specific to Neo4j and a policy graph needs to be constructed in advance. The enforcement ap-proach in XACML4G can work with property-graph compatible datastores, whereas GQRA is considered for Cypher-based datastores.

The current work has highlighted further questions concerning the differences and use cases of the two categories for access control approaches. For example, the *filtered results* category is typically used in retrieving authorized resources based on the defined policy. The *permit-deny access* category is less commonly used in access control for databases, but still considered in applications to allow authorized users to perform specific actions (e.g., edit or delete) on particular resources and vice versa. Furthermore, we plan to add more aspects in our comparison and include other works focusing on protecting property graph-structured data that have been excluded due to coarse-grained access control or a lack of application.

# ACKNOWLEDGEMENTS

5To be added later (because of double blind review). This work has been partly supported by the LIT Secure and Correct Systems Lab funded by the State of Upper Austria and the Linz Institute of Technology. This work has also been supported by the COMET-K2 Center of the Linz Center of Mechatronics (LCM) funded by the Austrian federal government and the federal state of Upper Austria.